\documentclass[aps,prl,twocolumn,superscriptaddress,amsmath,amssymb]{revtex4-2}

\input{definitions}

\begin{document}

\pacs{\input{pacs}}

\title{A test of light-lepton universality in the rates of inclusive semileptonic \B-meson decays at \belletwo}

\author{L.~Aggarwal\,\orcidlink{0000-0002-0909-7537}} 
\author{H.~Ahmed\,\orcidlink{0000-0003-3976-7498}} 
\author{H.~Aihara\,\orcidlink{0000-0002-1907-5964}} 
\author{N.~Akopov\,\orcidlink{0000-0002-4425-2096}} 
\author{A.~Aloisio\,\orcidlink{0000-0002-3883-6693}} 
\author{N.~Anh~Ky\,\orcidlink{0000-0003-0471-197X}} 
\author{D.~M.~Asner\,\orcidlink{0000-0002-1586-5790}} 
\author{H.~Atmacan\,\orcidlink{0000-0003-2435-501X}} 
\author{T.~Aushev\,\orcidlink{0000-0002-6347-7055}} 
\author{V.~Aushev\,\orcidlink{0000-0002-8588-5308}} 
% ...
\collaboration{The Belle II Collaboration}

\begin{abstract}
We present the first measurement of the ratio of branching fractions of inclusive semileptonic \B-meson decays, $\RXemu=\BFXenu/\BFXmunu$, a precision test of electron-muon universality, 
using data corresponding to \lumion from electron-positron collisions collected with the \belletwo detector. In events where the partner \B meson is fully reconstructed, we use fits to the lepton momentum spectra above $1.3\,\gevc$ to obtain $\RXemu = 1.007 \pm 0.009~(\mathrm{stat}) \pm 0.019~(\mathrm{syst})$, which is the most precise lepton-universality test of its kind and agrees with the standard-model expectation. 
\end{abstract}

\maketitle

In the standard model, all charged leptons share the same electroweak coupling, a symmetry known as lepton universality. Semileptonic \B-meson decays involving the quark transition $b \to c \,(e, \mu, \tau) \,\nu$ provide excellent sensitivity to potential lepton-universality-violating (LUV) physics. Persistent evidence for LUV in the rate of semileptonic decays to $\tau$ leptons relative to the light leptons, $\ell \in (e, \mu)$, has been found in the combination of results from the BaBar, Belle, and LHCb experiments~\cite{babar_1, babar_2, belle_hadronic, belle_semileptonic, belle_polarization, lhcb_1, lhcb_2, lhcb_3}.
Previous direct searches for LUV between the light leptons in semileptonic \B-meson decays have measured the branching-fraction ratio in a single exclusive charmed hadron decay mode~\cite{Remu} or the shapes of kinematic distributions of all decays to charmed hadrons~\cite{rayvt}. 

We present here the first measurement of the inclusive branching-fraction ratio $\RXemu=\BFXenu/\BFXmunu$, the most precise test of $e-\mu$ universality in semileptonic \B-meson decays to date. We indicate with $X$ the generic hadronic final state of the semileptonic decay of any flavor of \B meson originating from \btoclnu or, rarely, \btoulnu quark transitions. 
We use a \belletwo collision data set collected between 2019 and 2021 at a center-of-mass (c.m.)\ energy of $\sqrt{s}=10.58$~GeV, corresponding to the mass of the \FourS resonance, which almost exclusively decays into a pair of \B mesons of opposite flavor (\BzBzb, \BpBm). The integrated luminosity of the data set is \lumion, equivalent to approximately $198\times 10^6$ \BB pairs. We use an additional off-resonance collision data set, collected at an energy $\SI{60}{\MeV}$ below the \FourS resonance and corresponding to an integrated luminosity of \lumioff, to determine expected backgrounds from continuum processes $\epem\to\qqbar$, where $q$ indicates $u, d, s$, or $c$ quarks.
We reconstruct \B mesons decaying fully hadronically (the partner \B) and associate remaining particles with their accompanying \B meson (the signal \B). We identify leptons from among these remaining particles and extract the signal yield from a fit to the distribution of \pell, the lepton momentum in the rest frame of the signal \B meson.

The \belletwo detector~\cite{b2tdr, b2tip} operates at the SuperKEKB asymmetric-energy electron-positron collider~\cite{superkekb} at KEK. The detector consists of several nested subsystems arranged in a closed cylinder around the interaction region and nearly coaxial with the beams. The cylindrical portion is referred to as the barrel, which is closed by the forward and backward endcaps. The innermost subsystem is the vertex detector, composed of two layers of silicon pixels and four outer layers of silicon strip detectors. During data collection for this analysis the outermost pixel layer only covered \SI{15}{\percent} of the azimuthal angle. Charged-particle trajectories (tracks) are reconstructed by a small-cell drift chamber (CDC) filled with a He (\SI{50}{\percent}) and $\mathrm{C}_2\mathrm{H}_6$ (\SI{50}{\percent}) gas mixture, which also provides a measurement of ionization energy loss for particle identification. A Cherenkov-light imaging and time-of-propagation detector provides charged pion and kaon identification in the barrel region, while in the forward endcap a proximity-focusing, ring-imaging Cherenkov detector with an aerogel radiator is used.
An electromagnetic calorimeter (ECL) consisting of Cs(I) crystals provides photon and electron identification in the barrel and both endcaps. 
All of the above subsystems are embedded in a uniform 1.5~T magnetic field that is nearly aligned with the electron beam and is generated by a superconducting solenoid situated outside the ECL.
The outermost subsystem, the \KL and muon identification detector, consists of scintillator strips in the endcaps and the inner part of the barrel, and resistive plate chambers in the outer barrel, interleaved with iron plates that serve as a magnetic flux return yoke.

We use Monte Carlo simulation to produce signal and background models, and to calculate reconstruction efficiencies and detector acceptance. The processes are simulated with the \texttt{EvtGen}~\cite{evtgen}, \texttt{PYTHIA}~\cite{pythia8}, and \texttt{KKMC}~\cite{kkmc} software packages. Final-state radiation of photons from stable charged particles is simulated using the \texttt{PHOTOS} software package~\cite{PHOTOS}. Detector simulation is performed with the \texttt{GEANT4}~\cite{AGOSTINELLI2003250} software package.  Simulated beam-induced backgrounds are added to the events~\cite{BeamBKG}. Events are subsequently reconstructed and analyzed in the same fashion as the collision data with the Belle~II analysis software framework, \texttt{basf2}~\cite{basf2, basf2-zenodo}.
The simulated $\epem \to \FourS \to \BB$ samples contain known semileptonic and hadronic \B decays. 
The signal model includes the following known exclusive decays (charge conjugation is implied throughout): \bdlnu, \bdslnu, and \bddslnu, where \Ddoublestar collectively indicates the excited charmed states \Dzstar, \Doneprime, \Done, and \Dtwostar, whose masses and widths are taken from Ref.~\cite{pdg}. The \bdanddslnu decays are modeled with the Boyd-Grinstein-Lebed~\cite{BGL, DlnuBGL, DstlnuBGL} form-factor parametrization. The modeling of \bddslnu decays is based on the Bernlochner-Ligeti-Robinson (BLR) model~\cite{bernlochner_dstst, bernlochner_rdstst}.

Semileptonic \B decays into the nonresonant final states \bdanddspipilnu and \bdanddsetalnu are used to model the difference between the sum of individual branching fractions of exclusive decays, \bdanddslnu and \bddslnu, and the measured total semileptonic \B decay width~\cite{pdg}. These ``gap modes" are included in dedicated simulated samples that use intermediate, broad \Ddoublestar resonances and are modeled with BLR. Resonant and nonresonant Cabibbo-suppressed semileptonic \B decays including \btoulnu quark transitions are simulated simultaneously in a hybrid model~\cite{xu_hybrid} that is updated according to Ref.~\cite{pdg2016}.

We use the Full Event Interpretation (\FEI) algorithm~\cite{fei} to reconstruct the signal \B meson's partner in a fully hadronic decay mode, labeled \btag. This separates the final-state particles from the two \B mesons as well as increases the signal purity. 
We use three variables to maximize the purity of the \btag selection: the beam-constrained mass $\mbc=\sqrt{\left(\sqrt{s}/2\right)^{2}-|{\vec{p}_{B}}^{\,*}|^{2}}$, the energy difference $\deltae = E^{*}_{B} - \sqrt{s}/2$, and a confidence score produced by the \FEI to classify \B mesons, \sigprob, which has a value between zero (indicating low confidence in the tag reconstruction) and one (high confidence). The quantities ${\vec{p}_{B}}^{\,*}$ and $E^{*}_{B}$ denote the \btag momentum vector and energy in the c.m.\ frame, respectively.
We select \btag candidates with $\mbc \in[5.2725, 5.2850]\gevcc$, $\deltae \in[-0.15, 0.10]\gev$ and $\sigprob>0.1$. If multiple \btag candidates pass these selections in an event, we choose the one with the highest value of \sigprob. Approximately $82\%$ of the selected tag candidates are correctly reconstructed, accounting for roughly $0.1\%$ of all $\Upsilon(4S)$ events~\cite{fei}.

We reconstruct signal-lepton candidates with ${\pell>1.3\,\gevc}$ from the remaining tracks after the \btag reconstruction. We require that the lepton charge corresponds to the charge of a primary lepton from the semileptonic decay of a signal $B$ meson that has the opposite flavor to the \btag candidate. The lepton track candidates are extrapolated to the point of closest approach to the measured interaction point, which is required to be within $\SI{1}{\centi\meter}$ in the radial direction and within $\SI{3}{\centi\meter}$ along the beam axis, and are required to point within the CDC angular acceptance.

Muon candidates are required to have transverse momentum ${p_T>0.4\,\gevc}$ and are identified by means of a discriminator defined as the ratio 
$\mathcal{L}_{\mu}/\left(\mathcal{L}_{e}+\mathcal{L}_{\mu}+\mathcal{L}_{\pi}+\mathcal{L}_{K}+\mathcal{L}_{p}+\mathcal{L}_{d}\right)$, where the identification likelihood $\mathcal{L}_i$ for each charged-particle hypothesis $i$ combines particle-identification information from all subdetectors that provide it.  The resulting efficiency is measured from dedicated control channels to be on average $90\%$ for $p>1\,\gevc$, corresponding to an average muon misidentification probability for pions and kaons of $3\%$.

Electron candidates are required to have ${p_T>0.3\,\gevc}$. We correct their four-momenta to recover bremsstrahlung radiation by adding energy depositions in the ECL (clusters) that are not matched to any track and that are found within a cone centered on the electron direction. The opening angle of this cone depends on the momentum magnitude and is optimized using simulation. We validate the bremsstrahlung correction using an inclusive sample of $\jpsi\to e^+e^-$ candidates in experimental data.
Electron candidates are identified by means of a multiclass boosted-decision-tree classifier that exploits several ECL-cluster observables in combination with particle-identification likelihoods from the other \belletwo subsystems~\cite{LeptonIDBDT} defined analogously to the muon likelihood. The classifier thresholds are tuned in a three-dimensional grid of lab-frame momentum (\plab), polar angle (\thetalab), and charge ($q$) intervals to achieve a uniform 80\% identification efficiency. The misidentification probability for pions (kaons) with $p>1\,\gevc$ in the barrel is on average ${0.01(<0.001)\%}$.
If two or more signal-lepton candidates from the same event pass the above selections, we select the lepton with the highest identification likelihood.
We obtain correction weights, typically near 1.0, and uncertainties, for lepton-identification efficiencies and hadron-misidentification probabilities from auxiliary measurements in discrete intervals of (\plab, \thetalab, $q$) using dedicated data samples.
We calibrate the lepton-identification efficiencies using \jpsill, \llgamma, and \eell events. For charged kaons, we calibrate misidentification probabilities using \dstardkpipi events; for charged pions, we use \kspipi and \taupair events, where in the latter case one $\tau$ lepton is reconstructed in its decay modes with three charged hadrons.

All remaining ECL clusters not associated with a track that pass the following quality criteria are then combined to form the $X$ system. ECL clusters are required to be more than $\SI{30}{\centi\meter}$ away from the nearest extrapolated track and to have energies greater than $0.04$, $0.055$, and $\SI{0.09}{\gev}$ in the forward, barrel, and backward regions of the ECL, respectively. 
Tracks are required to be consistent with originating from the interaction point (within $\SI{2}{\centi\meter}$ in the radial direction and $\SI{4}{\centi\meter}$ along the beam axis), be in the CDC polar-angle acceptance, and have at least one measurement point in the CDC. Mass hypotheses are assigned to each charged particle by checking particle-identification criteria in a specific sequence (electron, muon, kaon, proton) and assigning the hypothesis of the first satisfied criterion. Remaining charged particles are considered to be pions.  

We suppress continuum background with a boosted decision tree trained on simulated data that exploits 21 event-topology variables built from particle candidates that pass the same selection criteria as those used for the $X$ system reconstruction. These variables quantify the spatial distribution of momentum and energy in the events in order to discriminate between \BB events, which are largely isotropic, and continuum events, which tend to have a back-to-back structure~\cite{ref:physics_of_the_B_factories}. We select events identified as \BB-like, which rejects $55\%$ of the continuum background while retaining $97\%$ of the \BB candidates. 

We describe the remaining continuum background using off-resonance data with the yield scaled by the squared ratio of off- to on-resonance c.m.\ collision energies ${\scaleoffres=(\ecmsoffres/\ecmsonres)^2=0.989}$ to account for a factor of $1/s$ in the $\epem\to q\bar{q}$ cross-section. The energy and momentum of particles in this data set are also scaled by $1/\sqrt{\scaleoffres}=1.006$ to account for the reduced c.m.\ energy available with respect to collisions at the \FourS resonance.

We extract the signal yields, $N^\text{meas}_{\ell}$, with simultaneous binned maximum-likelihood template fits to the $p_e^B$ and $p_{\mu}^B$ spectra in the range $\pell\in[1.3, 2.3]\gevc$ subdivided into 10 equal intervals (bins), where the last bin includes any overflow events. The lower limit on \pell is chosen to reduce backgrounds, and to suppress \BtoXtaunu decays to a negligible level. 

We define three components for each lepton flavor, which we fit simultaneously. The signal component, \BtoXellnu, has an unconstrained yield. The continuum component has a Gaussian constraint on its yield derived from off-resonance data. The background component mostly contains events with hadrons misidentified as leptons (fakes) and correctly reconstructed lepton candidates originating mainly from decays of charmed hadrons (secondaries). The yield of this component has a Gaussian constraint derived from a fit to data in a same-charge control channel containing events with two \B mesons reconstructed with the same flavor and therefore enriched with fakes and secondaries, but also including \BtoXellnu from neutral $B$-meson oscillations. We perform this control-channel fit with electrons and muons simultaneously and with unconstrained background and \BtoXellnu yields. We further verify that it is robust against arbitrary variations of the predicted yields of any of the components.

The statistical and systematic uncertainties are incorporated in the likelihood definition via nuisance parameters, one for each \pell bin for each component. Constraints on the nuisance parameters are encoded in a global covariance matrix for bins and components, constructed by summing the covariance matrices of all individual uncertainty sources. 

The uncertainties associated with the lepton-identification-efficiency and hadron-misidentification weights are provided by auxiliary measurements, as previously described. They are propagated to \RXemu uncertainties under the following assumptions for leptons (or hadron fakes) of a given type: uncertainties within the same (\plab, \thetalab, $q$) bin are fully correlated for events from different components; statistical (systematic) uncertainties are fully uncorrelated (correlated) for events in different bins and components.

We obtain event weights from branching-fraction uncertainties by performing Gaussian variations of each using central values and widths from the best experimental determinations and their uncertainties. For \BtoXuellnu and \bdanddslnu, we use the latest values~\cite{pdg, hflav}, combining the results of neutral and charged \B mesons under the assumption of isospin symmetry in the latter case. For the remaining \btoclnu decays, not all possible final states have been measured to date. We estimate their unknown branching fractions by extrapolating from existing measurements to the unobserved \Ddoublestar final-state decays, again assuming isospin symmetry.
Among the nonresonant gap modes, only the decay $B\rightarrow D^{(*)}\,\pi^+\,\pi^-\,\ell\,\bar{\nu}_\ell$ is measured~\cite{dpipi_babar}. This result is extrapolated to the other charge configurations to estimate their total branching fractions. The remaining gap modes, \bdanddsetalnu, are assigned a \SI{100}{\percent} branching fraction uncertainty. The fit to data reduces the uncertainties on the gap-mode branching fractions by exploiting the differences in shape between these modes and the remaining signal.
Form-factor parameters are varied within their uncertainties (including correlations) using the HAMMER software package~\cite{hammer}. Uncertainties in the number of selected signal events from uncertainties in branching fractions and form-factor parameters are assumed to be fully correlated between the electron and muon channels. 

The ratios of the \btag reconstruction efficiencies in data and simulation for each used \B hadronic decay mode are all compatible between the electron and muon channels within their statistical uncertainties.
Therefore, we conclude they fully cancel out in the \RXemu ratio, and assign no further systematic uncertainty.

After all selections and corrections, we determine total signal efficiencies by extracting the selected signal yields, $N^{\mathrm{sel}}_{\ell}$, from fits to the simulated spectra and dividing by the number of generated events in the full phase space, $N^\text{gen}_{\ell}$. The electron efficiency is ${(1.77\pm0.04)\times10^{-3}}$, and the muon efficiency is ${(2.14\pm0.06)\times10^{-3}}$. The correlation between the two efficiencies due to shared systematic uncertainties is $0.76$.

\begin{figure*}[]
	\centering
	\includegraphics[width=0.49\linewidth]{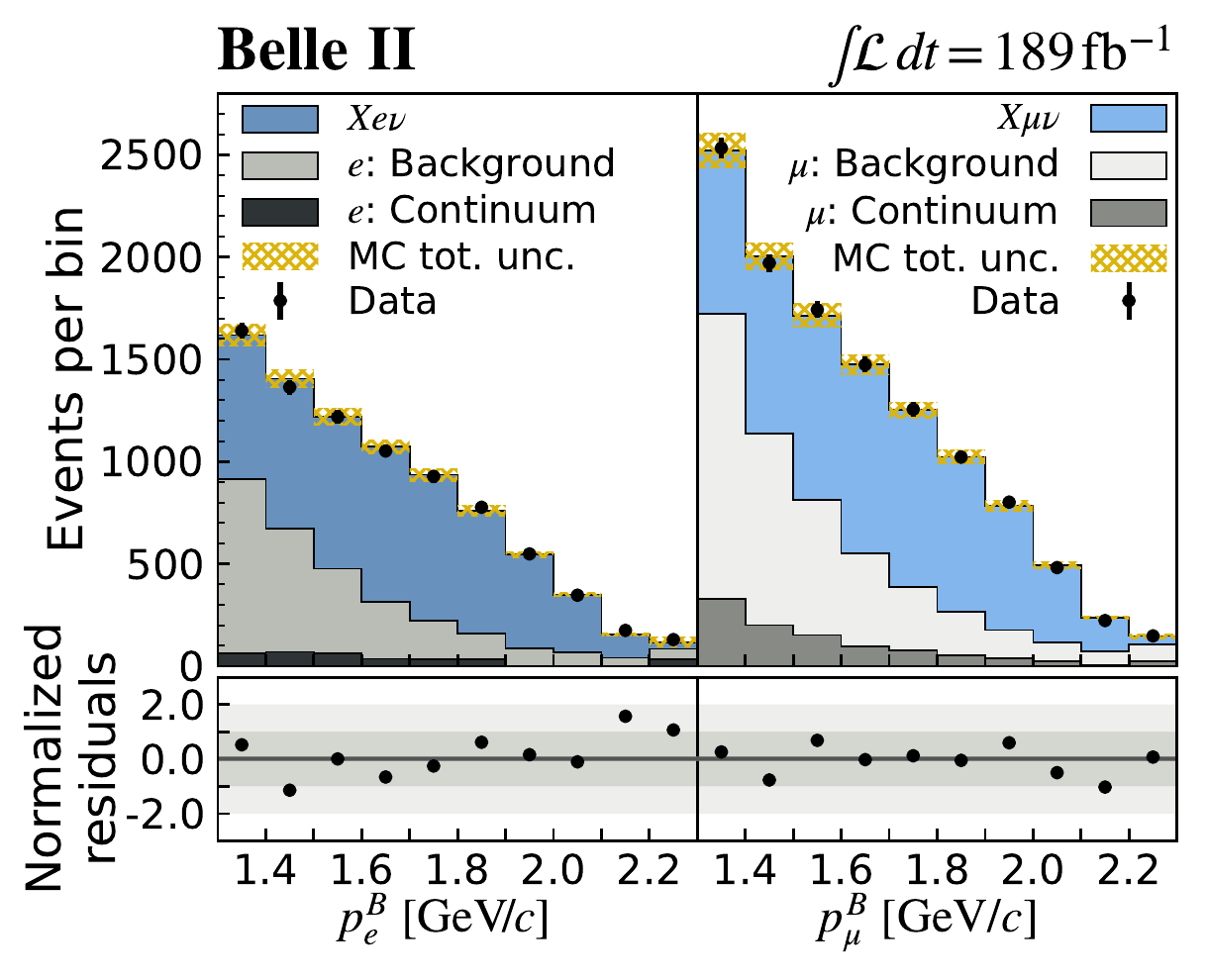}
	\includegraphics[width=0.49\linewidth]{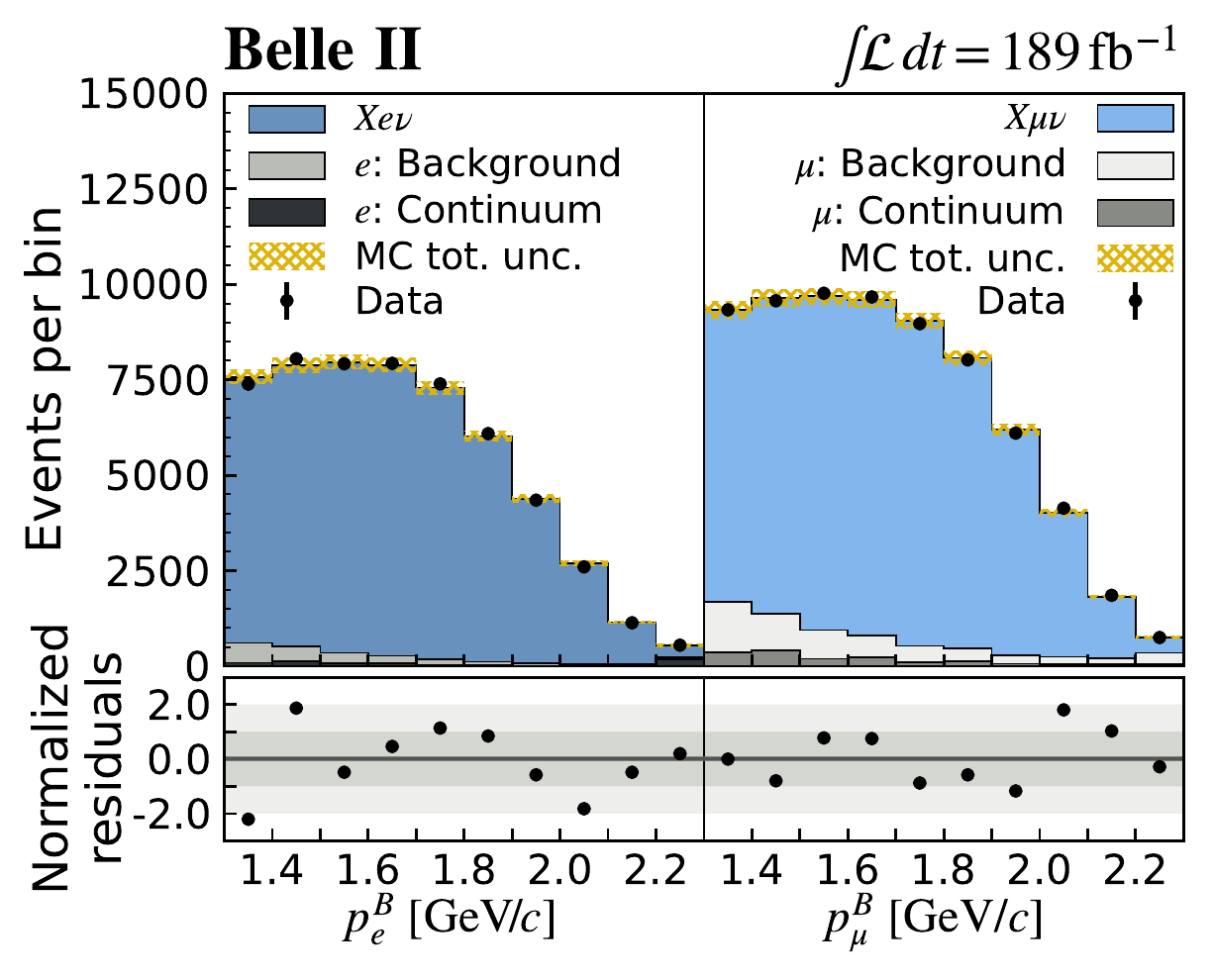}
	\caption{Same-charge control channel (left) and opposite-charge signal (right) spectra of the lepton momentum in the \bsig rest frame, \pell, with the fit results overlaid. The background component mostly contains events with fake or secondary leptons. The last bin contains overflow events. The hatched area shows the total statistical plus systematic uncertainty, added in quadrature in each bin.}
	\label{fig:fit_templates}
\end{figure*}

We fit the experimental \pell spectra in the same-charge control and opposite-charge signal samples, as shown in Fig.~\ref{fig:fit_templates}. We measure ${N_{e}^{\text{meas}}=50960\pm290}$ and ${N_{\mu}^{\text{meas}}=61300\pm400}$ signal events in the electron and muon modes, respectively, with a correlation of 0.027. The global $\chi^2$ value after the fit is $21.5$ with 18 degrees of freedom, corresponding to a $p$ value of 0.25.

From the yields, we calculate \RXemu using
\begin{equation}
	\label{eq:Remu_and_eff}
	\RXemu = \frac{N_{e}^{\text{meas}}}{N_{\mu}^{\text{meas}}}\cdot\frac{N_{\mu}^{\text{sel}}}{N_{e}^{\text{sel}}}\cdot\frac{N_{e}^{\text{gen}}}{N_{\mu}^{\text{gen}}}.
\end{equation}

We estimate the size of each systematic uncertainty by first fitting the simulated spectrum with only statistical fluctuations allowed. We then enable fluctuations from one systematic source and take the quadrature difference of the two to be the uncertainty from that source. We further validate these uncertainties by generating a large number of test data sets obtained by modifying the simulated data set, each corresponding to a specific systematic variation, and observing the resulting variation in the extracted value of \RXemu. The resulting uncertainties are summarized in Table~\ref{tab:uncertainties}. The largest uncertainty, of \SI{1.9}{\percent}, is associated with the lepton-identification efficiencies and misidentification probabilities. In the \RXemu ratio, branching-fraction and form-factor uncertainties largely cancel, with residual uncertainties arising from coupling between signal and background template shapes. Uncertainties associated with track finding efficiencies are negligible.

\begin{table}[]
	\caption{Statistical and systematic uncertainties on the value of \RXemu from the most significant sources.}	
	\begin{tabularx}{\columnwidth}{l @{\hskip 10mm} l}
		\hline 
		\hline
		Source                      & Uncertainty [\%] \\ 
		\hline
		Sample size                 & 0.9 \\   
		Lepton identification       & 1.9 \\
		\xlnu branching fractions  & 0.2 \\
		\xclnu form factors         & 0.1 \\
		\hline
		Total                       & 2.1 \\
		\hline
		\hline
	\end{tabularx}
	\label{tab:uncertainties}
\end{table}

We find an \RXemu value of
\begin{equation}\label{eq:full_phase_space_result}
	\RXemu = 1.007 \pm 0.009~(\mathrm{stat}) \pm 0.019~(\mathrm{syst}),
\end{equation}
which agrees with a previous measurement from Belle in exclusive $B\to D^* \ell \nu$ decays~\cite{Remu}.
In order to reduce model dependence, we also provide a fiducial measurement by recalculating $N^{\text{gen}}_{\ell}$ of Eq.~\eqref{eq:Remu_and_eff} in the restricted phase space defined by selecting events with a generated \B-frame lepton momentum above $1.3\,\gevc$, leading to an overall scaling of \RXemu by $0.998$. The result is
\begin{equation}
	\begin{split}
		\RXemufid &= 1.005 \pm 0.009~(\mathrm{stat}) \\
		&\pm 0.019~(\mathrm{syst}).      
	\end{split}
\end{equation}
In order to test the dependence of the result on the chosen lower threshold on \pell, we measure \RXemu while changing the nominal value of $1.3\,\gevc$ to $1.1$, $1.2$, and $1.4\,\gevc$. The values are mutually consistent with a $p$-value of 0.27, taking into account the correlations between uncertainties of the four measurements. Similarly, the result is consistent between subsets of the full data set when split by lepton charge, tag flavor, and by data-taking period. We find that the bremsstrahlung recovery procedure has negligible impact on the result. Furthermore, we check the impact on \RXemu of the modeling of charmed $D$ meson decays by varying the branching ratio of each decay $D\to K + \mathrm{anything}$ within its uncertainty as provided in Ref.~\cite{pdg} while fixing the total event normalization. The effect is negligible. No evidence for a significant bias associated with the selection of a single candidate in the case of multiple candidates as described in Ref.~\cite{bcs_bias} is observed.

Our result is the most precise branching fraction-based test of electron-muon universality in semileptonic \B decays. The measurement in the full phase space, Eq.~\ref{eq:full_phase_space_result}, is consistent with the standard-model prediction of $1.006 \pm 0.001$~\cite{kvos}.

This work, based on data collected using the Belle II detector, which was built and commissioned prior to March 2019, was supported by
Science Committee of the Republic of Armenia Grant No.~20TTCG-1C010;
Australian Research Council and research Grants
No.~DE220100462,
No.~DP180102629,
No.~DP170102389,
No.~DP170102204,
No.~DP150103061,
No.~FT130100303,
No.~FT130100018,
and
No.~FT120100745;
Austrian Federal Ministry of Education, Science and Research,
Austrian Science Fund
No.~P~31361-N36
and
No.~J4625-N,
and
Horizon 2020 ERC Starting Grant No.~947006 ``InterLeptons'';
Natural Sciences and Engineering Research Council of Canada, Compute Canada and CANARIE;
Chinese Academy of Sciences and research Grant No.~QYZDJ-SSW-SLH011,
National Natural Science Foundation of China and research Grants
No.~11521505,
No.~11575017,
No.~11675166,
No.~11761141009,
No.~11705209,
and
No.~11975076,
LiaoNing Revitalization Talents Program under Contract No.~XLYC1807135,
Shanghai Pujiang Program under Grant No.~18PJ1401000,
Shandong Provincial Natural Science Foundation Project~ZR2022JQ02,
and the CAS Center for Excellence in Particle Physics (CCEPP);
the Ministry of Education, Youth, and Sports of the Czech Republic under Contract No.~LTT17020 and
Charles University Grant No.~SVV 260448 and
the Czech Science Foundation Grant No.~22-18469S;
European Research Council, Seventh Framework PIEF-GA-2013-622527,
Horizon 2020 ERC-Advanced Grants No.~267104 and No.~884719,
Horizon 2020 ERC-Consolidator Grant No.~819127,
Horizon 2020 Marie Sklodowska-Curie Grant Agreement No.~700525 "NIOBE"
and
No.~101026516,
and
Horizon 2020 Marie Sklodowska-Curie RISE project JENNIFER2 Grant Agreement No.~822070 (European grants);
L'Institut National de Physique Nucl\'{e}aire et de Physique des Particules (IN2P3) du CNRS (France);
BMBF, DFG, HGF, MPG, and AvH Foundation (Germany);
Department of Atomic Energy under Project Identification No.~RTI 4002 and Department of Science and Technology (India);
Israel Science Foundation Grant No.~2476/17,
U.S.-Israel Binational Science Foundation Grant No.~2016113, and
Israel Ministry of Science Grant No.~3-16543;
Istituto Nazionale di Fisica Nucleare and the research grants BELLE2;
Japan Society for the Promotion of Science, Grant-in-Aid for Scientific Research Grants
No.~16H03968,
No.~16H03993,
No.~16H06492,
No.~16K05323,
No.~17H01133,
No.~17H05405,
No.~18K03621,
No.~18H03710,
No.~18H05226,
No.~19H00682, 
No.~22H00144,
No.~26220706,
and
No.~26400255,
the National Institute of Informatics, and Science Information NETwork 5 (SINET5), 
and
the Ministry of Education, Culture, Sports, Science, and Technology (MEXT) of Japan;  
National Research Foundation (NRF) of Korea Grants
No.~2016R1\-D1A1B\-02012900,
No.~2018R1\-A2B\-3003643,
No.~2018R1\-A6A1A\-06024970,
No.~2018R1\-D1A1B\-07047294,
No.~2019R1\-I1A3A\-01058933,
No.~2022R1\-A2C\-1003993,
and
No.~RS-2022-00197659,
Radiation Science Research Institute,
Foreign Large-size Research Facility Application Supporting project,
the Global Science Experimental Data Hub Center of the Korea Institute of Science and Technology Information
and
KREONET/GLORIAD;
Universiti Malaya RU grant, Akademi Sains Malaysia, and Ministry of Education Malaysia;
Frontiers of Science Program Contracts
No.~FOINS-296,
No.~CB-221329,
No.~CB-236394,
No.~CB-254409,
and
No.~CB-180023, and No.~SEP-CINVESTAV research Grant No.~237 (Mexico);
the Polish Ministry of Science and Higher Education and the National Science Center;
the Ministry of Science and Higher Education of the Russian Federation,
Agreement No.~14.W03.31.0026, and
the HSE University Basic Research Program, Moscow;
University of Tabuk research Grants
No.~S-0256-1438 and No.~S-0280-1439 (Saudi Arabia);
Slovenian Research Agency and research Grants
No.~J1-9124
and
No.~P1-0135;
Agencia Estatal de Investigacion, Spain
Grant No.~RYC2020-029875-I
and
Generalitat Valenciana, Spain
Grant No.~CIDEGENT/2018/020
Ministry of Science and Technology and research Grants
No.~MOST106-2112-M-002-005-MY3
and
No.~MOST107-2119-M-002-035-MY3,
and the Ministry of Education (Taiwan);
Thailand Center of Excellence in Physics;
TUBITAK ULAKBIM (Turkey);
National Research Foundation of Ukraine, project No.~2020.02/0257,
and
Ministry of Education and Science of Ukraine;
the U.S. National Science Foundation and research Grants
No.~PHY-1913789 
and
No.~PHY-2111604, 
and the U.S. Department of Energy and research Awards
No.~DE-AC06-76RLO1830, 
No.~DE-SC0007983, 
No.~DE-SC0009824, 
No.~DE-SC0009973, 
No.~DE-SC0010007, 
No.~DE-SC0010073, 
No.~DE-SC0010118, 
No.~DE-SC0010504, 
No.~DE-SC0011784, 
No.~DE-SC0012704, 
No.~DE-SC0019230, 
No.~DE-SC0021274, 
No.~DE-SC0022350; 
and
the Vietnam Academy of Science and Technology (VAST) under Grant No.~DL0000.05/21-23.

These acknowledgements are not to be interpreted as an endorsement of any statement made
by any of our institutes, funding agencies, governments, or their representatives.

We thank the SuperKEKB team for delivering high-luminosity collisions;
the KEK cryogenics group for the efficient operation of the detector solenoid magnet;
the KEK computer group and the NII for on-site computing support and SINET6 network support;
and the raw-data centers at BNL, DESY, GridKa, IN2P3, INFN, and the University of Victoria for offsite computing support.

\bibliographystyle{apsrev4-2}
\bibliography{references}

\begin{thebibliography}{37}%
\makeatletter
\providecommand \@ifxundefined [1]{%
 \@ifx{#1\undefined}
}%
\providecommand \@ifnum [1]{%
 \ifnum #1\expandafter \@firstoftwo
 \else \expandafter \@secondoftwo
 \fi
}%
\providecommand \@ifx [1]{%
 \ifx #1\expandafter \@firstoftwo
 \else \expandafter \@secondoftwo
 \fi
}%
\providecommand \natexlab [1]{#1}%
\providecommand \enquote  [1]{``#1''}%
\providecommand \bibnamefont  [1]{#1}%
\providecommand \bibfnamefont [1]{#1}%
\providecommand \citenamefont [1]{#1}%
\providecommand \href@noop [0]{\@secondoftwo}%
\providecommand \href [0]{\begingroup \@sanitize@url \@href}%
\providecommand \@href[1]{\@@startlink{#1}\@@href}%
\providecommand \@@href[1]{\endgroup#1\@@endlink}%
\providecommand \@sanitize@url [0]{\catcode `\\12\catcode `\$12\catcode
  `\&12\catcode `\#12\catcode `\^12\catcode `\_12\catcode `\%12\relax}%
\providecommand \@@startlink[1]{}%
\providecommand \@@endlink[0]{}%
\providecommand \url  [0]{\begingroup\@sanitize@url \@url }%
\providecommand \@url [1]{\endgroup\@href {#1}{\urlprefix }}%
\providecommand \urlprefix  [0]{URL }%
\providecommand \Eprint [0]{\href }%
\providecommand \doibase [0]{https://doi.org/}%
\providecommand \selectlanguage [0]{\@gobble}%
\providecommand \bibinfo  [0]{\@secondoftwo}%
\providecommand \bibfield  [0]{\@secondoftwo}%
\providecommand \translation [1]{[#1]}%
\providecommand \BibitemOpen [0]{}%
\providecommand \bibitemStop [0]{}%
\providecommand \bibitemNoStop [0]{.\EOS\space}%
\providecommand \EOS [0]{\spacefactor3000\relax}%
\providecommand \BibitemShut  [1]{\csname bibitem#1\endcsname}%
\let\auto@bib@innerbib\@empty
\bibitem [{\citenamefont {Lees}\ \emph {et~al.}(2012)\citenamefont {Lees} \emph
  {et~al.}}]{babar_1}%
  \BibitemOpen
  \bibfield  {author} {\bibinfo {author} {\bibfnamefont {J.~P.}\ \bibnamefont
  {Lees}} \emph {et~al.} (\bibinfo {collaboration} {BaBar Collaboration}),\
  }\href {https://link.aps.org/doi/10.1103/PhysRevLett.109.101802} {\bibfield
  {journal} {\bibinfo  {journal} {Phys. Rev. Lett.}\ }\textbf {\bibinfo
  {volume} {109}},\ \bibinfo {pages} {101802} (\bibinfo {year}
  {2012})}\BibitemShut {NoStop}%
\bibitem [{\citenamefont {Lees}\ \emph {et~al.}(2013)\citenamefont {Lees} \emph
  {et~al.}}]{babar_2}%
  \BibitemOpen
  \bibfield  {author} {\bibinfo {author} {\bibfnamefont {J.~P.}\ \bibnamefont
  {Lees}} \emph {et~al.} (\bibinfo {collaboration} {BaBar Collaboration}),\
  }\href {https://link.aps.org/doi/10.1103/PhysRevD.88.072012} {\bibfield
  {journal} {\bibinfo  {journal} {Phys. Rev. D}\ }\textbf {\bibinfo {volume}
  {88}},\ \bibinfo {pages} {072012} (\bibinfo {year} {2013})}\BibitemShut
  {NoStop}%
\bibitem [{\citenamefont {Huschle}\ \emph {et~al.}(2015)\citenamefont {Huschle}
  \emph {et~al.}}]{belle_hadronic}%
  \BibitemOpen
  \bibfield  {author} {\bibinfo {author} {\bibfnamefont {M.}~\bibnamefont
  {Huschle}} \emph {et~al.} (\bibinfo {collaboration} {Belle Collaboration}),\
  }\href {https://link.aps.org/doi/10.1103/PhysRevD.92.072014} {\bibfield
  {journal} {\bibinfo  {journal} {Phys. Rev. D}\ }\textbf {\bibinfo {volume}
  {92}},\ \bibinfo {pages} {072014} (\bibinfo {year} {2015})}\BibitemShut
  {NoStop}%
\bibitem [{\citenamefont {Caria}\ \emph {et~al.}(2020)\citenamefont {Caria}
  \emph {et~al.}}]{belle_semileptonic}%
  \BibitemOpen
  \bibfield  {author} {\bibinfo {author} {\bibfnamefont {G.}~\bibnamefont
  {Caria}} \emph {et~al.} (\bibinfo {collaboration} {Belle Collaboration}),\
  }\href {https://doi.org/10.1103/PhysRevLett.124.161803} {\bibfield  {journal}
  {\bibinfo  {journal} {Phys. Rev. Lett.}\ }\textbf {\bibinfo {volume} {124}},\
  \bibinfo {pages} {161803} (\bibinfo {year} {2020})}\BibitemShut {NoStop}%
\bibitem [{\citenamefont {Hirose}\ \emph {et~al.}(2017)\citenamefont {Hirose}
  \emph {et~al.}}]{belle_polarization}%
  \BibitemOpen
  \bibfield  {author} {\bibinfo {author} {\bibfnamefont {S.}~\bibnamefont
  {Hirose}} \emph {et~al.} (\bibinfo {collaboration} {Belle Collaboration}),\
  }\href {https://link.aps.org/doi/10.1103/PhysRevLett.118.211801} {\bibfield
  {journal} {\bibinfo  {journal} {Phys. Rev. Lett.}\ }\textbf {\bibinfo
  {volume} {118}},\ \bibinfo {pages} {211801} (\bibinfo {year}
  {2017})}\BibitemShut {NoStop}%
\bibitem [{\citenamefont {Aaij}\ \emph {et~al.}(2015)\citenamefont {Aaij} \emph
  {et~al.}}]{lhcb_1}%
  \BibitemOpen
  \bibfield  {author} {\bibinfo {author} {\bibfnamefont {R.}~\bibnamefont
  {Aaij}} \emph {et~al.} (\bibinfo {collaboration} {LHCb Collaboration}),\
  }\href {https://link.aps.org/doi/10.1103/PhysRevLett.115.111803} {\bibfield
  {journal} {\bibinfo  {journal} {Phys. Rev. Lett.}\ }\textbf {\bibinfo
  {volume} {115}},\ \bibinfo {pages} {111803} (\bibinfo {year} {2015})},\
  \bibinfo {note} {[Erratum: Phys. Rev. Lett. 115, 159901 (2015)]}\BibitemShut
  {NoStop}%
\bibitem [{\citenamefont {Aaij}\ \emph
  {et~al.}(2018{\natexlab{a}})\citenamefont {Aaij} \emph {et~al.}}]{lhcb_2}%
  \BibitemOpen
  \bibfield  {author} {\bibinfo {author} {\bibfnamefont {R.}~\bibnamefont
  {Aaij}} \emph {et~al.} (\bibinfo {collaboration} {LHCb Collaboration}),\
  }\href {https://link.aps.org/doi/10.1103/PhysRevLett.120.171802} {\bibfield
  {journal} {\bibinfo  {journal} {Phys. Rev. Lett.}\ }\textbf {\bibinfo
  {volume} {120}},\ \bibinfo {pages} {171802} (\bibinfo {year}
  {2018}{\natexlab{a}})}\BibitemShut {NoStop}%
\bibitem [{\citenamefont {Aaij}\ \emph
  {et~al.}(2018{\natexlab{b}})\citenamefont {Aaij} \emph {et~al.}}]{lhcb_3}%
  \BibitemOpen
  \bibfield  {author} {\bibinfo {author} {\bibfnamefont {R.}~\bibnamefont
  {Aaij}} \emph {et~al.} (\bibinfo {collaboration} {LHCb Collaboration}),\
  }\href {https://link.aps.org/doi/10.1103/PhysRevD.97.072013} {\bibfield
  {journal} {\bibinfo  {journal} {Phys. Rev. D}\ }\textbf {\bibinfo {volume}
  {97}},\ \bibinfo {pages} {072013} (\bibinfo {year}
  {2018}{\natexlab{b}})}\BibitemShut {NoStop}%
\bibitem [{\citenamefont {Waheed}\ \emph {et~al.}(2019)\citenamefont {Waheed}
  \emph {et~al.}}]{Remu}%
  \BibitemOpen
  \bibfield  {author} {\bibinfo {author} {\bibfnamefont {E.}~\bibnamefont
  {Waheed}} \emph {et~al.} (\bibinfo {collaboration} {Belle Collaboration}),\
  }\href {https://link.aps.org/doi/10.1103/PhysRevD.100.052007} {\bibfield
  {journal} {\bibinfo  {journal} {Phys. Rev. D}\ }\textbf {\bibinfo {volume}
  {100}},\ \bibinfo {pages} {052007} (\bibinfo {year} {2019})}\BibitemShut
  {NoStop}%
\bibitem [{\citenamefont {van Tonder}\ \emph {et~al.}(2021)\citenamefont {van
  Tonder} \emph {et~al.}}]{rayvt}%
  \BibitemOpen
  \bibfield  {author} {\bibinfo {author} {\bibfnamefont {R.}~\bibnamefont {van
  Tonder}} \emph {et~al.} (\bibinfo {collaboration} {Belle Collaboration}),\
  }\href {https://link.aps.org/doi/10.1103/PhysRevD.104.112011} {\bibfield
  {journal} {\bibinfo  {journal} {Phys. Rev. D}\ }\textbf {\bibinfo {volume}
  {104}},\ \bibinfo {pages} {112011} (\bibinfo {year} {2021})}\BibitemShut
  {NoStop}%
\bibitem [{\citenamefont {Abe}\ \emph {et~al.}(2010)\citenamefont {Abe} \emph
  {et~al.}}]{b2tdr}%
  \BibitemOpen
  \bibfield  {author} {\bibinfo {author} {\bibfnamefont {T.}~\bibnamefont
  {Abe}} \emph {et~al.} (\bibinfo {collaboration} {Belle II Collaboration}),\
  }\href@noop {} {\  (\bibinfo {year} {2010})},\ \Eprint
  {https://arxiv.org/abs/1011.0352} {arXiv:1011.0352} \BibitemShut {NoStop}%
\bibitem [{\citenamefont {Altmannshofer}\ \emph {et~al.}(2019)\citenamefont
  {Altmannshofer} \emph {et~al.}}]{b2tip}%
  \BibitemOpen
  \bibfield  {author} {\bibinfo {author} {\bibfnamefont {W.}~\bibnamefont
  {Altmannshofer}} \emph {et~al.},\ }\href
  {https://doi.org/10.1093/ptep/ptz106} {\bibfield  {journal} {\bibinfo
  {journal} {Prog. Theor. Exp. Phys.}\ }\textbf {\bibinfo {volume} {2019}},\
  \bibinfo {pages} {123C01} (\bibinfo {year} {2019})},\ \bibinfo {note}
  {123C01}\BibitemShut {NoStop}%
\bibitem [{\citenamefont {Akai}\ \emph {et~al.}(2018)\citenamefont {Akai} \emph
  {et~al.}}]{superkekb}%
  \BibitemOpen
  \bibfield  {author} {\bibinfo {author} {\bibfnamefont {K.}~\bibnamefont
  {Akai}} \emph {et~al.} (\bibinfo {collaboration} {SuperKEKB accelerator
  team}),\ }\href {https://doi.org/10.1016/j.nima.2018.08.017} {\bibfield
  {journal} {\bibinfo  {journal} {Nucl. Instrum. Meth. A}\ }\textbf {\bibinfo
  {volume} {907}},\ \bibinfo {pages} {188} (\bibinfo {year}
  {2018})}\BibitemShut {NoStop}%
\bibitem [{\citenamefont {Lange}(2001)}]{evtgen}%
  \BibitemOpen
  \bibfield  {author} {\bibinfo {author} {\bibfnamefont {D.~J.}\ \bibnamefont
  {Lange}},\ }\href {https://doi.org/10.1016/S0168-9002(01)00089-4} {\bibfield
  {journal} {\bibinfo  {journal} {Nucl. Instrum. Meth. A}\ }\textbf {\bibinfo
  {volume} {462}},\ \bibinfo {pages} {152} (\bibinfo {year}
  {2001})}\BibitemShut {NoStop}%
\bibitem [{\citenamefont {Sjöstrand}\ \emph {et~al.}(2015)\citenamefont
  {Sjöstrand} \emph {et~al.}}]{pythia8}%
  \BibitemOpen
  \bibfield  {author} {\bibinfo {author} {\bibfnamefont {T.}~\bibnamefont
  {Sjöstrand}} \emph {et~al.},\ }\href
  {https://doi.org/10.1016/j.cpc.2015.01.024} {\bibfield  {journal} {\bibinfo
  {journal} {Comput. Phys. Commun.}\ }\textbf {\bibinfo {volume} {191}},\
  \bibinfo {pages} {159} (\bibinfo {year} {2015})}\BibitemShut {NoStop}%
\bibitem [{\citenamefont {Jadach}\ \emph {et~al.}(2000)\citenamefont {Jadach}
  \emph {et~al.}}]{kkmc}%
  \BibitemOpen
  \bibfield  {author} {\bibinfo {author} {\bibfnamefont {S.}~\bibnamefont
  {Jadach}} \emph {et~al.},\ }\href
  {https://doi.org/10.1016/S0010-4655(00)00048-5} {\bibfield  {journal}
  {\bibinfo  {journal} {Comput. Phys. Commun.}\ }\textbf {\bibinfo {volume}
  {130}},\ \bibinfo {pages} {260} (\bibinfo {year} {2000})}\BibitemShut
  {NoStop}%
\bibitem [{\citenamefont {Barberio}\ \emph {et~al.}(1991)\citenamefont
  {Barberio}, \citenamefont {{van Eijk}},\ and\ \citenamefont {Was}}]{PHOTOS}%
  \BibitemOpen
  \bibfield  {author} {\bibinfo {author} {\bibfnamefont {E.}~\bibnamefont
  {Barberio}}, \bibinfo {author} {\bibfnamefont {B.}~\bibnamefont {{van
  Eijk}}},\ and\ \bibinfo {author} {\bibfnamefont {Z.}~\bibnamefont {Was}},\
  }\href {https://www.sciencedirect.com/science/article/pii/001046559190012A}
  {\bibfield  {journal} {\bibinfo  {journal} {Comput. Phys. Commun.}\ }\textbf
  {\bibinfo {volume} {66}},\ \bibinfo {pages} {115} (\bibinfo {year}
  {1991})}\BibitemShut {NoStop}%
\bibitem [{\citenamefont {Agostinelli}\ \emph {et~al.}(2003)\citenamefont
  {Agostinelli} \emph {et~al.}}]{AGOSTINELLI2003250}%
  \BibitemOpen
  \bibfield  {author} {\bibinfo {author} {\bibfnamefont {S.}~\bibnamefont
  {Agostinelli}} \emph {et~al.} (\bibinfo {collaboration} {\textsc{Geant4}
  Collaboration}),\ }\href {https://doi.org/10.1016/S0168-9002(03)01368-8}
  {\bibfield  {journal} {\bibinfo  {journal} {Nucl. Instrum. Meth. A}\ }\textbf
  {\bibinfo {volume} {506}},\ \bibinfo {pages} {250} (\bibinfo {year}
  {2003})}\BibitemShut {NoStop}%
\bibitem [{\citenamefont {Natochii}\ \emph {et~al.}(2022)\citenamefont
  {Natochii} \emph {et~al.}}]{BeamBKG}%
  \BibitemOpen
  \bibfield  {author} {\bibinfo {author} {\bibfnamefont {A.}~\bibnamefont
  {Natochii}} \emph {et~al.},\ }\href@noop {} {\  (\bibinfo {year} {2022})},\
  \Eprint {https://arxiv.org/abs/2203.05731} {arXiv:2203.05731} \BibitemShut
  {NoStop}%
\bibitem [{\citenamefont {Kuhr}\ \emph {et~al.}(2019)\citenamefont {Kuhr} \emph
  {et~al.}}]{basf2}%
  \BibitemOpen
  \bibfield  {author} {\bibinfo {author} {\bibfnamefont {T.}~\bibnamefont
  {Kuhr}} \emph {et~al.} (\bibinfo {collaboration} {Belle II Framework Software
  Group}),\ }\href {https://doi.org/10.1007/s41781-018-0017-9} {\bibfield
  {journal} {\bibinfo  {journal} {Comput. Softw. Big Sci.}\ }\textbf {\bibinfo
  {volume} {3}},\ \bibinfo {pages} {1} (\bibinfo {year} {2019})}\BibitemShut
  {NoStop}%
\bibitem [{bas()}]{basf2-zenodo}%
  \BibitemOpen
  \href {https://doi.org/10.5281/zenodo.5574115} {\bibinfo {title} {{Belle II
  Analysis Software Framework (basf2)}}},\ \bibinfo {howpublished}
  {\url{https://doi.org/10.5281/zenodo.5574115}}\BibitemShut {NoStop}%
\bibitem [{\citenamefont {Workman}\ \emph {et~al.}(2022)\citenamefont {Workman}
  \emph {et~al.}}]{pdg}%
  \BibitemOpen
  \bibfield  {author} {\bibinfo {author} {\bibfnamefont {R.~L.}\ \bibnamefont
  {Workman}} \emph {et~al.} (\bibinfo {collaboration} {Particle Data Group}),\
  }\href {https://doi.org/10.1093/ptep/ptac097} {\bibfield  {journal} {\bibinfo
   {journal} {PTEP}\ }\textbf {\bibinfo {volume} {2022}},\ \bibinfo {pages}
  {083C01} (\bibinfo {year} {2022})}\BibitemShut {NoStop}%
\bibitem [{\citenamefont {Boyd}\ \emph {et~al.}(1995)\citenamefont {Boyd},
  \citenamefont {Grinstein},\ and\ \citenamefont {Lebed}}]{BGL}%
  \BibitemOpen
  \bibfield  {author} {\bibinfo {author} {\bibfnamefont {C.~G.}\ \bibnamefont
  {Boyd}}, \bibinfo {author} {\bibfnamefont {B.}~\bibnamefont {Grinstein}},\
  and\ \bibinfo {author} {\bibfnamefont {R.~F.}\ \bibnamefont {Lebed}},\ }\href
  {https://doi.org/10.1103/PhysRevLett.74.4603} {\bibfield  {journal} {\bibinfo
   {journal} {Phys. Rev. Lett.}\ }\textbf {\bibinfo {volume} {74}},\ \bibinfo
  {pages} {4603} (\bibinfo {year} {1995})}\BibitemShut {NoStop}%
\bibitem [{\citenamefont {Glattauer}\ \emph {et~al.}(2016)\citenamefont
  {Glattauer} \emph {et~al.}}]{DlnuBGL}%
  \BibitemOpen
  \bibfield  {author} {\bibinfo {author} {\bibfnamefont {R.}~\bibnamefont
  {Glattauer}} \emph {et~al.} (\bibinfo {collaboration} {Belle
  Collaboration}),\ }\href {https://doi.org/10.1103/PhysRevD.93.032006}
  {\bibfield  {journal} {\bibinfo  {journal} {Phys. Rev. D}\ }\textbf {\bibinfo
  {volume} {93}},\ \bibinfo {pages} {032006} (\bibinfo {year}
  {2016})}\BibitemShut {NoStop}%
\bibitem [{\citenamefont {Ferlewicz}\ \emph {et~al.}(2021)\citenamefont
  {Ferlewicz}, \citenamefont {Urquijo},\ and\ \citenamefont
  {Waheed}}]{DstlnuBGL}%
  \BibitemOpen
  \bibfield  {author} {\bibinfo {author} {\bibfnamefont {D.}~\bibnamefont
  {Ferlewicz}}, \bibinfo {author} {\bibfnamefont {P.}~\bibnamefont {Urquijo}},\
  and\ \bibinfo {author} {\bibfnamefont {E.}~\bibnamefont {Waheed}},\ }\href
  {https://link.aps.org/doi/10.1103/PhysRevD.103.073005} {\bibfield  {journal}
  {\bibinfo  {journal} {Phys. Rev. D}\ }\textbf {\bibinfo {volume} {103}},\
  \bibinfo {pages} {073005} (\bibinfo {year} {2021})}\BibitemShut {NoStop}%
\bibitem [{\citenamefont {Bernlochner}\ \emph {et~al.}(2018)\citenamefont
  {Bernlochner}, \citenamefont {Ligeti},\ and\ \citenamefont
  {Robinson}}]{bernlochner_dstst}%
  \BibitemOpen
  \bibfield  {author} {\bibinfo {author} {\bibfnamefont {F.~U.}\ \bibnamefont
  {Bernlochner}}, \bibinfo {author} {\bibfnamefont {Z.}~\bibnamefont
  {Ligeti}},\ and\ \bibinfo {author} {\bibfnamefont {D.~J.}\ \bibnamefont
  {Robinson}},\ }\href {https://doi.org/10.1103/PhysRevD.97.075011} {\bibfield
  {journal} {\bibinfo  {journal} {Phys. Rev. D}\ }\textbf {\bibinfo {volume}
  {97}},\ \bibinfo {pages} {075011} (\bibinfo {year} {2018})}\BibitemShut
  {NoStop}%
\bibitem [{\citenamefont {Bernlochner}\ and\ \citenamefont
  {Ligeti}(2017)}]{bernlochner_rdstst}%
  \BibitemOpen
  \bibfield  {author} {\bibinfo {author} {\bibfnamefont {F.~U.}\ \bibnamefont
  {Bernlochner}}\ and\ \bibinfo {author} {\bibfnamefont {Z.}~\bibnamefont
  {Ligeti}},\ }\href {https://doi.org/10.1103/PhysRevD.95.014022} {\bibfield
  {journal} {\bibinfo  {journal} {Phys. Rev. D}\ }\textbf {\bibinfo {volume}
  {95}},\ \bibinfo {pages} {014022} (\bibinfo {year} {2017})}\BibitemShut
  {NoStop}%
\bibitem [{\citenamefont {Ramirez}\ \emph {et~al.}(1990)\citenamefont
  {Ramirez}, \citenamefont {Donoghue},\ and\ \citenamefont
  {Burdman}}]{xu_hybrid}%
  \BibitemOpen
  \bibfield  {author} {\bibinfo {author} {\bibfnamefont {C.}~\bibnamefont
  {Ramirez}}, \bibinfo {author} {\bibfnamefont {J.~F.}\ \bibnamefont
  {Donoghue}},\ and\ \bibinfo {author} {\bibfnamefont {G.}~\bibnamefont
  {Burdman}},\ }\href {https://doi.org/10.1103/PhysRevD.41.1496} {\bibfield
  {journal} {\bibinfo  {journal} {Phys. Rev. D}\ }\textbf {\bibinfo {volume}
  {41}},\ \bibinfo {pages} {1496} (\bibinfo {year} {1990})}\BibitemShut
  {NoStop}%
\bibitem [{\citenamefont {Patrignani}\ \emph {et~al.}(2016)\citenamefont
  {Patrignani} \emph {et~al.}}]{pdg2016}%
  \BibitemOpen
  \bibfield  {author} {\bibinfo {author} {\bibfnamefont {C.}~\bibnamefont
  {Patrignani}} \emph {et~al.} (\bibinfo {collaboration} {Particle Data
  Group}),\ }\href {https://doi.org/10.1088/1674-1137/40/10/100001} {\bibfield
  {journal} {\bibinfo  {journal} {Chinese Physics C}\ }\textbf {\bibinfo
  {volume} {40}},\ \bibinfo {pages} {100001} (\bibinfo {year}
  {2016})}\BibitemShut {NoStop}%
\bibitem [{\citenamefont {Keck}\ \emph {et~al.}(2019)\citenamefont {Keck} \emph
  {et~al.}}]{fei}%
  \BibitemOpen
  \bibfield  {author} {\bibinfo {author} {\bibfnamefont {T.}~\bibnamefont
  {Keck}} \emph {et~al.},\ }\href {https://doi.org/10.1007/s41781-019-0021-8}
  {\bibfield  {journal} {\bibinfo  {journal} {Comput. Softw. Big Sci.}\
  }\textbf {\bibinfo {volume} {3}},\ \bibinfo {pages} {6} (\bibinfo {year}
  {2019})}\BibitemShut {NoStop}%
\bibitem [{\citenamefont {Milesi}\ \emph {et~al.}(2020)\citenamefont {Milesi},
  \citenamefont {Tan},\ and\ \citenamefont {Urquijo}}]{LeptonIDBDT}%
  \BibitemOpen
  \bibfield  {author} {\bibinfo {author} {\bibfnamefont {M.}~\bibnamefont
  {Milesi}}, \bibinfo {author} {\bibfnamefont {J.}~\bibnamefont {Tan}},\ and\
  \bibinfo {author} {\bibfnamefont {P.}~\bibnamefont {Urquijo}},\ }\href
  {https://doi.org/10.1051/epjconf/202024506023} {\bibfield  {journal}
  {\bibinfo  {journal} {EPJ Web Conf.}\ }\textbf {\bibinfo {volume} {245}},\
  \bibinfo {pages} {06023} (\bibinfo {year} {2020})}\BibitemShut {NoStop}%
\bibitem [{\citenamefont {{Ed.\ A.~J.~Bevan, B.~Golob, Th.~Mannel, S.~Prell,
  and B.~D.~Yabsley}}(2014)}]{ref:physics_of_the_B_factories}%
  \BibitemOpen
  \bibfield  {author} {\bibinfo {author} {\bibnamefont {{Ed.\ A.~J.~Bevan,
  B.~Golob, Th.~Mannel, S.~Prell, and B.~D.~Yabsley}}},\ }\href
  {https://doi.org/10.1140/epjc/s10052-014-3026-9} {\bibfield  {journal}
  {\bibinfo  {journal} {Eur. Phys. J. C}\ }\textbf {\bibinfo {volume} {74}}
  (\bibinfo {year} {2014})},\ \bibinfo {note} {p.~109}\BibitemShut {NoStop}%
\bibitem [{\citenamefont {Amhis}\ \emph {et~al.}(2021)\citenamefont {Amhis}
  \emph {et~al.}}]{hflav}%
  \BibitemOpen
  \bibfield  {author} {\bibinfo {author} {\bibfnamefont {Y.}~\bibnamefont
  {Amhis}} \emph {et~al.} (\bibinfo {collaboration} {Heavy Flavor Averaging
  Group (HFLAV)}),\ }\href {https://doi.org/10.1140%2Fepjc%2Fs10052-020-8156-7}
  {\bibfield  {journal} {\bibinfo  {journal} {Eur. Phys. J. C}\ }\textbf
  {\bibinfo {volume} {81}} (\bibinfo {year} {2021})}\BibitemShut {NoStop}%
\bibitem [{\citenamefont {Lees}\ \emph {et~al.}(2016)\citenamefont {Lees} \emph
  {et~al.}}]{dpipi_babar}%
  \BibitemOpen
  \bibfield  {author} {\bibinfo {author} {\bibfnamefont {J.~P.}\ \bibnamefont
  {Lees}} \emph {et~al.} (\bibinfo {collaboration} {BABAR Collaboration}),\
  }\href {https://doi.org/10.1103/PhysRevLett.116.041801} {\bibfield  {journal}
  {\bibinfo  {journal} {Phys. Rev. Lett.}\ }\textbf {\bibinfo {volume} {116}},\
  \bibinfo {pages} {041801} (\bibinfo {year} {2016})}\BibitemShut {NoStop}%
\bibitem [{\citenamefont {Bernlochner}\ \emph {et~al.}(2020)\citenamefont
  {Bernlochner} \emph {et~al.}}]{hammer}%
  \BibitemOpen
  \bibfield  {author} {\bibinfo {author} {\bibfnamefont {F.~U.}\ \bibnamefont
  {Bernlochner}} \emph {et~al.},\ }\href
  {https://doi.org/10.1140%2Fepjc%2Fs10052-020-8304-0} {\bibfield  {journal}
  {\bibinfo  {journal} {Eur. Phys. J. C}\ }\textbf {\bibinfo {volume} {80}}
  (\bibinfo {year} {2020})}\BibitemShut {NoStop}%
\bibitem [{\citenamefont {Koppenburg}(2019)}]{bcs_bias}%
  \BibitemOpen
  \bibfield  {author} {\bibinfo {author} {\bibfnamefont {P.}~\bibnamefont
  {Koppenburg}},\ }\href@noop {} {\  (\bibinfo {year} {2019})},\ \Eprint
  {https://arxiv.org/abs/1703.01128} {arXiv:1703.01128 [hep-ex]} \BibitemShut
  {NoStop}%
\bibitem [{\citenamefont {Rahimi}\ and\ \citenamefont {Vos}(2022)}]{kvos}%
  \BibitemOpen
  \bibfield  {author} {\bibinfo {author} {\bibfnamefont {M.}~\bibnamefont
  {Rahimi}}\ and\ \bibinfo {author} {\bibfnamefont {K.~K.}\ \bibnamefont
  {Vos}},\ }\href {https://doi.org/10.1007/JHEP11(2022)007} {\bibfield
  {journal} {\bibinfo  {journal} {J. High Energ. Phys.}\ }\textbf {\bibinfo
  {volume} {11}},\ \bibinfo {pages} {007 (2022)}}\BibitemShut {NoStop}%
\end{thebibliography}%

\end{document}